\documentclass{article}

\usepackage{amsmath, amssymb}

\def \Rm#1{\mbox{\rm #1}}
\def \lsem      {\raise1pt\hbox{\Rm {[\kern-.12em[}}}
\def \rsem      {\raise1pt\hbox{\Rm {]\kern-.12em]}}}
\def \sem#1{\mbox{\lsem$#1$\rsem}}

\def\>{\ensuremath{\rangle}}
\def\<{\ensuremath{\langle}}
\def\qzz{\ensuremath{|0\>_q\<0|}}
\def\qoo{\ensuremath{|1\>_q\<1|}}
\def\qzo{\ensuremath{|0\>_q\<1|}}
\def\qoz{\ensuremath{|1\>_q\<0|}}
\def\h{\ensuremath{\mathcal{H}}}
\def\lh{\ensuremath{\mathcal{L(H)}}}
\def\dh{\ensuremath{\mathcal{DH}}}
\def\qh{\ensuremath{\mathcal{QH}}}
\def\ph{\ensuremath{\mathcal{PH}}}
\def\th{\ensuremath{\mathcal{TH}}}
\def\t{\ensuremath{\mathcal{T}}}

\def\z{\ensuremath{{\bf 0}}}

\def\e{\ensuremath{\mathcal{E}}}
\def\f{\ensuremath{\mathcal{F}}}

\def\le{\ensuremath{\sqsubseteq}}
\def\ge{\ensuremath{\sqsupseteq}}

\newcommand {\true} {\mbox{\bf{true}}}
\newcommand {\abort} {\mbox{\bf{abort}}}
\newcommand {\sskip} {\mbox{\bf{skip}}}
\newcommand {\measure} {\mbox{\bf{measure}}}
\newcommand {\then} {\mbox{\bf{then}}}
\newcommand {\eelse} {\mbox{\bf{else}}}
\newcommand {\while} {\mbox{\bf{while}}}
\newcommand {\ddo} {\mbox{\bf{do}}}
\newcommand{\tr}{{\rm Tr}}

\newtheorem{thm}{Theorem}[section]
\newtheorem{cor}[thm]{Corollary}
\newtheorem{lem}[thm]{Lemma}

\newtheorem{defn}[thm]{Definition}
\begin{document}

\title{Proof rules for purely quantum programs}
\author{Yuan Feng,\ Runyao Duan,\ Zhengfeng Ji,\ and\ Mingsheng\
Ying\\ \\
\small State Key Laboratory of Intelligent Technology and Systems,\\
\small Department of Computer Science and Technology,\\
\small Tsinghua University, Beijing, China, 100084}

\maketitle \thispagestyle{empty}

\begin{abstract}
We apply the notion of quantum predicate proposed by D'Hondt and
Panangaden to analyze a purely quantum language fragment which
describes the quantum part of a future quantum computer in Knill's
architecture. The denotational semantics, weakest precondition
semantics, and weakest liberal precondition semantics of this
language fragment are introduced. To help reasoning about quantum
programs involving quantum loops, we extend proof rules for
classical probabilistic programs to our purely quantum programs.

\end{abstract}

\section{Introduction}

\nocite{To03,To04,FD05,Yi03}

The theory of quantum computing has attracted considerable research
efforts in the past twenty years. Benefiting from the possibility of
superposition of different states and the linearity of quantum
operations, quantum computing may provide considerable speedup over
its classical analogue \cite{Sh94,Gr96,Gr97}. The existing quantum
algorithms, however, are described at a very low level: they are
usually represented as quantum circuits. A few works have been done
in developing quantum programming languages which identify and
promote high-level abstractions. Knill \cite{Kn96} moved the first
step by outlining a set of basic principles for writing quantum
pseudo-code; while the first actual quantum programming language is
due to \"{O}mer \cite{Om98}. After that, Sanders and Zuliani
\cite{SZ00}, Bettelli et al. \cite{BCS03}, and Selinger \cite{Se04}
also proposed various quantum languages each having different
features.

The standard weakest precondition calculus \cite{Di76} and its
probabilistic extension \cite{MMS96} have been successful in
reasoning about the correctness and even the rigorous derivation of
classical programs. This success motivates us to develop analogous
tools for quantum programs. Sanders and Zuliani \cite{SZ00} have
provided for their qGCL a stepwise refinement mechanics. The
approach, however, is classical in the sense that they treated
quantum programs as special cases of probabilistic programs. As a
consequence, known results about probabilistic weakest precondition
calculus can be applied directly to quantum programs. Indeed, Butler
and Hartel \cite{BH99} have used it to reason about Grover's
algorithm.

The first step towards really $quantum$ weakest precondition
calculus was made by D'Hondt and Panangaden \cite{HP05}. They
proposed the brilliant idea that we can treat an observable,
mathematically described by a Hermitian matrix, as the quantum
analogue of `predicate'. The elegant duality between
state-transformer semantics and the weakest precondition semantics
of quantum programs was then proven to hold in a more direct way.

In this paper, we apply the ideas in \cite{HP05} to analyze a purely
quantum language fragment describing the quantum part of a potential
quantum computer in Knill's architecture \cite{Kn96}. The syntax
follows Selinger's style but we consider only purely quantum data.
We introduce the denotational semantics for this purely quantum
language fragment, which are represented by super-operators. The
weakest precondition semantics corresponding to total correctness
and weakest liberal precondition semantics corresponding to partial
correctness are also introduced. To help reasoning about quantum
loops, we also extend proof rules for classical probabilistic
programs to our purely quantum programs.

\section{Preliminaries}
In this section, we review some notions and results from \cite{HP05}
which are the basis of our work.

Let $\h$ be the associated Hilbert space of the quantum system we
are concerned with, and $\lh$ the set of linear operators (or
complex matrices, we do not distinguish between these two notions)
on $\h$. Let $\dh$ be the set of all density operators on $\h$, that
is,
$$\dh:=\{\ \rho\in\lh\ |\ \z\le \rho,\ \tr(\rho)\leq
\mbox{1}\},$$ where $\z$ denotes the zero operator. The convention
of allowing the trace of a density matrix to be less than 1 makes it
possible to represent both the actual state (by the normalized
density matrix) and the probability with which the state is reached
(by the trace of the density matrix) \cite{Se04}. The partial order
$\le$ is defined on the set of all matrices with the same dimension
by letting $M\le N$ if $N-M$ is positive. Then the set of quantum
programs over $\h$ is defined as
$$\qh:=\{\e\in\dh\rightarrow
\dh\ |\ \e\mbox{ is a super-operator}\}.$$ We lift the partial order
in $\dh$ to the one in $\qh$ by letting $\e \sqsubseteq\f$ if
$\e(\rho)\leq \f(\rho)$  for any $\rho\in\dh$. It is proved in
\cite{HP05} that the two sets $\dh$ and $\qh$ are both CPOs.

In D'Hondt and Panangaden's approach, a quantum predicate is
described by a Hermitian positive matrix with the maximum eigenvalue
bounded by 1. To be specific, the set of quantum predicates on
Hilbert space $\h$ is defined by
$$\ph:=\{M\in\lh\ |\ M^\dag=M,\  \mbox{\z\ \le M\ \le\ I}\}.$$
For any $\rho\in\dh$ and $M\in\ph$, the degree of $\rho$ satisfying
$M$ is denoted by the expression $\tr M\rho$. It is exactly the
expectation of the outcomes when performing a measurement
represented by $M$ on the state $\rho$.

The `healthy' predicate transformers which exactly characterize all
valid quantum programs are proved to be those who are linear and
completely positive \cite{HP05}. That is, there exists an isomorphic
map between the set of healthy quantum predicate transformers
$$\th:=\{\t\in\ph\rightarrow\ph\ |\ \t \mbox{ is linear and completely positve}\}$$
and the set of quantum programs $\qh$ defined above, just as the
cases for classical standard \cite{Di76} and probabilistic programs
\cite{MMS96}.

\section{The purely quantum language fragment and its semantics}

In this section, we concentrate our attention on the purely quantum
fragment of a general programming language. That is, only quantum
data but no classical data are considered. Following Knill's QRAM
model \cite{Kn96}, a quantum computer in the future possibly
consists of a general-purpose classical computer which controls a
special quantum hardware device. The quantum device contains a
large, but finite number of individually addressable quantum bits.
The classical controller communicates with the quantum device by
sending a sequence of control instructions and receiving the results
of the measurements on quantum bits. Our purely quantum language
considered here then aims at describing the action of the special
quantum device, rather than the behavior of the whole computer
including the classical controller.

Suppose $S,S_0$ and $S_1$ denote purely quantum programs,
$q_1,\dots,q_n$ and $q$ denote qubit-typed variables, and $U$
denotes a unitary transformation which applies on a
$2^n$-dimensional Hilbert space. The syntax of our purely quantum
language is defined as follows:
\begin{eqnarray*} S&::= &\ \abort\ |\ \sskip\ |\ q:=0\ |\
q_1,q_2,\dots,q_n*=U\ |\ S_0;S_1\ |\\
& &\ \measure\ q\ \then\ S_0\ \eelse\ S_1\ |\ \while\ q\ \ddo\ S.
\end{eqnarray*}

Here we borrow the notations from \cite{Se04} except for the loop
statements in which loop conditions are also purely quantum.
Intuitively, the statement $q:=0$ initializes qubit $q$ by setting
it to the standard state $|0\>$. Note that it is the only assignment
in the language. This is also why our language is functional rather
than imperative. The statement $q_1,q_2,\dots,q_n*=U$ applies the
unitary transformation $U$ on $n$ distinct qubits
$q_1,q_2,\dots,q_n$. We put the constraint that $q_1,q_2,\dots,q_n$
must be distinct to avoid syntactically some no-go operations such
as quantum cloning. The statement $\measure\ q\ \then\ S_0\ \eelse\
S_1$ first applies a measurement on qubit $q$, then executes $S_0$
or $S_1$ depending on whether the measurement result is $0$ or $1$.
The loop statement $\while\ q\ \ddo\ S$ measures qubit $q$ first. If
the result is 1, then it terminates; otherwise it executes $S$ and
the loop repeats.

Formally, we have the following denotational semantics:
\begin{defn}\label{def:opsemantics} For any purely quantum program
$S$, the denotational semantics of $S$ is a map $\sem{S}$ from $\dh$
to $\dh$ defined inductively in Figure \ref{tb:ope}, where $(\while\
q\ \ddo\ S)^0 :=\abort$ and $$(\while\ q\ \ddo\ S)^{i+1} :=\measure\
q\ \then\ S; (\while\ q\ \ddo\ S)^i\ \eelse\ \sskip.$$
\end{defn}

\begin{figure}[t] \caption{Denotational semantics}
\begin{eqnarray*}
\hline\\
\sem{\abort}\rho&:=& \z\\
\sem{\sskip}\rho&:=&\rho\\
\sem{q:=0}\rho&:=&\qzz \rho\qzz+\qzo \rho\qoz\\
\sem{\bar{q}*=U}\rho&:=&U_{\bar{q}} \rho U_{\bar{q}}^\dag\\
\sem{S_1;S_2}\rho&:=&\sem{S_2}(\sem{S_1}\rho)\\
\sem{\measure\ q\ \then\ S_0\ \eelse\ S_1}\rho&:=&\sem{S_0}(\qzz \rho \qzz) + \sem{S_1}(\qoo \rho\qoo)\\
\sem{\while\ q\ \ddo\ S}\rho&:=&\sqcup_{i=0}^\infty \sem{(\while\ q\
\ddo\ S)^i}\rho\\ \\\hline
\end{eqnarray*} \label{tb:ope}
\end{figure}

In Definition \ref{def:opsemantics}, $\bar{q}$ denotes the
abbreviation of $q_1,\dots,q_n$, $U_{\bar{q}}$ means applying $U$ on
the Hilbert space spanned by qubits $\bar{q}$, and $\qzz \rho\qzz$
denotes the application $|0\>\<0| \bullet |0\>\<0|$ on qubit $q$
when the initial state is $\rho$, leaving other qubits unchanged.
That is,
$$\qzz \rho\qzz=(I_{\h_1}\otimes |0\>\<0|\otimes I_{\h_2})\rho (I_{\h_1}\otimes |0\>\<0|\otimes
I_{\h_2})$$ for some appropriate Hilbert spaces $\h_1$ and $\h_2$.
In Section 4, we often omit the subscript $q$ for simplicity when no
confusion arises.

The following lemma shows that the denotational semantics of our
purely quantum programs are all super-operators. So they can be
physically implemented in a future quantum computer.

\begin{lem}\label{lem:opissuper}
For any purely quantum program $S$, the denotational semantics of
$S$ is a super-operator on $\dh$, i.e., $\sem{S}\in\qh$.
\end{lem}
{\it Proof.} We prove the theorem by induction on the structure of
$S$. When $S$ has the form other than quantum loop, the proof is
straightforward. So in what follows, we assume $S\equiv \while\ q\
\ddo\ S'$ and $\sem{S'}\in\qh$ for induction hypothesis.

To prove $\sem{S}\in\qh$, we need only to show that for any $i\geq
0$,
\begin{equation}\label{eq:inqh}
\sem{(\while\ q\ \ddo\ S)^i}\in\qh
\end{equation}
and
\begin{equation}\label{eq:defnloop}
\sem{(\while\ q\ \ddo\ S)^i}\le \sem{(\while\ q\ \ddo\ S)^{i+1}}.
\end{equation} Eq.(\ref{eq:inqh}) is easy to prove by induction on $i$.
To prove Eq.(\ref{eq:defnloop}), notice first that for any
$\rho\in\dh$, $\sem{\abort}\rho= \z$ is the bottom element of $\dh$.
So $\sem{\abort}$ is the bottom element of $\qh$ and then
Eq.(\ref{eq:defnloop}) holds trivially for the case $i=0$. Suppose
further Eq.(\ref{eq:defnloop}) holds for $i=k$. Then we calculate
that for any $\rho\in\dh$,
\begin{eqnarray*}
& &\sem{(\while\ q\ \ddo\ S)^{k+1}}\rho\\
&=&\sem{(\while\ q\ \ddo\
S)^{k}}(\sem{S}(\qzz \rho \qzz)) + \qoo \rho\qoo\hspace{3em}\mbox{by definition}\\
&\ge& \sem{(\while\ q\ \ddo\ S)^{k-1}}(\sem{S}(\qzz \rho \qzz)) +
\qoo \rho\qoo\\
& & \hspace{24em}\mbox{induction hypothesis}\\
&=&\sem{(\while\ q\ \ddo\ S)^{k}}\rho.\hspace{18.7em}\mbox{by
definition}
\end{eqnarray*}
Finally, from the fact that $\qh$ is a CPO we have $\sem{S}\in\qh$.
 \hfill $\square$

\vspace{1em}

Note that the syntax of the language we consider does not provide
the power to create new qubits. So by our purely quantum programs we
cannot implement all super-operators on $\dh$ since in general to
realize a super-operator we need to introduce some auxiliary qubits.
It seems to be a bad news. In practice, however, the number of
qubits a quantum program can use is restricted by the maximum a real
quantum computer can provide. The domain of the semantics of our
purely quantum programs is the Hilbert space associated with the
quantum device as a whole, so they indeed include all real
operations we can perform on a quantum computer.

Following the idea of quantum predicate presented in \cite{HP05}, we
define the weakest precondition semantics of our purely quantum
programs as follows:
\begin{defn}For any purely quantum program $S$, the
weakest precondition semantics of $S$ is defined by a map $wp.S$
from $\ph$ to $\ph$ defined inductively in Figure \ref{tb:ws}, where
$(\while\ q\ \ddo\ S)^i$ is defined in Definition
\ref{def:opsemantics}.
\end{defn}

\begin{figure}[t]\caption{weakest precondition semantics}
\begin{eqnarray*}
\hline \\
wp.\abort.M&:=&\z\\
wp.\sskip.M&:=&M\\
wp.(q:=0).M&:=&\qzz M\qzz+\qoz M\qzo\\
wp.(\bar{q}*=U).M&:=&U_{\bar{q}}^\dag MU_{\bar{q}}\\
wp.(S_1;S_2).M&:=&wp.S_1.(wp.S_2.M)\\
wp.(\measure\ q\ \then\ S_0\ \eelse\ S_1).M&:=&\sum_{i=0}^1|i\>_q\<i| wp.S_i.M|i\>_q\<i|\\
wp.(\while\ q\ \ddo\ S).M&:=&\sqcup_{i=0}^\infty wp.(\while\ q\
\ddo\ S)^i.M\\ \\ \hline
\end{eqnarray*}\label{tb:ws}
\end{figure}

An alternative definition of $wp.(\while\ q\ \ddo\ S).M$ is the
least fixed point $\mu X\cdot(\qzz wp.S.X \qzz + \qoo M\qoo)$. It is
easy to check that these two definitions are equivalent.

The following theorem shows a quantitative relation between
denotational semantics and weakest precondition semantics.
Intuitively, the expectation of observing any quantum predicate on
the output of a quantum program is equal to the expectation of
observing the weakest precondition of this predicate on the input
state.

\begin{thm}\label{thm:wpop}
For any purely quantum program $S$, quantum predicate $M\in\ph$, and
$\rho\in\dh$, we have
\begin{equation}\label{eq:wp}
\tr(wp.S.M)\rho = \tr M\sem{S}\rho
\end{equation}
\end{thm}
{\it Proof.} We need only to consider the case that $S\equiv \while\
q\ \ddo\ S'$ is a quantum loop. Other cases are easy to check.

Suppose Eq.(\ref{eq:wp}) holds for the program $S'$, i.e.,
\begin{equation}\label{hypothesis}
\forall M\in\ph; \rho\in\dh\cdot\tr(wp.S'.M)\rho=\tr M\sem{S'}\rho.
\end{equation}
We first prove by induction that for any $i\geq 0$
\begin{equation}\label{induction}
\forall M\in\ph; \rho\in\dh\cdot\tr(wp.S^i.M)\rho=\tr
M\sem{S^i}\rho.
\end{equation}

When $i=0$, Eq.(\ref{induction}) holds because both sides equal to
0. Suppose now that Eq.(\ref{induction}) holds for $i=k$. Then when
$i=k+1$, we calculate that for any $M\in\ph$ and $\rho\in\dh$,
\begin{eqnarray*}
& &\tr(wp.S^{k+1}.M\rho)\\
&=&\tr(\qzz wp.S'.(wp.S^k.M)\qzz + \qoo M\qoo)\rho\\
&=&\tr (wp.S'.(wp.S^k.M)\qzz\rho\qzz) + \tr M\qoo\rho\qoo\\
&=&\tr((wp.S^k.M)\sem{S'}\qzz\rho\qzz) + \tr M\qoo\rho\qoo\hspace{2.2em}\mbox{by Eq.(\ref{hypothesis})}\\
&=&\tr M\sem{S^k}(\sem{S'}\qzz \rho\qzz) + \tr M\qoo \rho\qoo\hspace{3.2em}\mbox{by induction hypothesis}\\
&=&\tr M\sem{S^{k+1}}\rho.
\end{eqnarray*}
So we deduce that Eq.(\ref{induction}) holds for any $i\geq 0$. And
then \begin{eqnarray*} \tr(wp.S.M)\rho &=&
\tr(\sqcup_i wp.S^i.M)\rho\\
&=&\sqcup_i \tr (wp.S^i.M)\rho\\
&=&\sqcup_i \tr M\sem{S^i}\rho\hspace{5em}\mbox{by Eq.(\ref{induction})}\\
&=& \tr M\sqcup_i \sem{S^i}\rho\\
&=& \tr M\sem{S}\rho.
\end{eqnarray*}
That completes our proof.
 \hfill $\square$

\vspace{1em}

 Taking $M=I$ in Eq.(\ref{eq:wp}), we have
$$\tr(wp.S.I)\rho = \tr\sem{S}\rho.$$ Notice that the righthand side
of the above equation denotes the probability the (un-normalized)
output state $\sem{S}\rho$ is reached. So intuitively, for any
purely quantum program $S$, the quantum predicate $wp.S.I$ denotes
the condition the program $S$ terminates, in analogy with the
predicate $wp.S.\true$ in classical standard setting and $wp.S.1$ in
classical probabilistic setting.

We have so far defined the weakest precondition semantics, which is
useful when we consider the total correctness of quantum programs.
That is, what we care is not only the correctness of the final state
when the program terminates, but also the condition and the
probability a quantum program can terminate. To deal with partial
correctness of quantum programs, we introduce the notion of weakest
liberal precondition semantics as follows:

\begin{defn}For any purely quantum program $S$, the weakest liberal precondition
semantics of $S$ is defined by a map $wlp.S$ from $\ph$ to $\ph$
defined inductively in Figure \ref{tb:wls}, where $(\while\ q\ \ddo\
S)^i$ is defined in Definition \ref{def:opsemantics}.
\end{defn}
\begin{figure}[t]\caption{weakest liberal precondition semantics}
\begin{eqnarray*}
 \hline\\
wlp.\abort.M&:=&I\\
wlp.\sskip.M&:=&M\\
wlp.(q:=0).M&:=&\qzz M\qzz+\qoz M\qzo\\
wlp.(\bar{q}*=U).M&:=&U_{\bar{q}}^\dag MU_{\bar{q}}\\
wlp.(S_1;S_2).M&:=&wlp.S_1(wlp.S_2.M)\\
wlp.(\measure\ q\ \then\ S_0\ \eelse\
S_1).M&:=&\sum_{i=0}^1|i\>_q\<i|wlp.S_i.M|i\>_q\<i|\\
wlp.(\while\ q\ \ddo\ S).M&:=&\sqcap_{i=0}^\infty wlp.(\while\ q\
\ddo\ S)^i.M
\\ \hline
\end{eqnarray*}\label{tb:wls}
\end{figure}

Analogous with weakest precondition semantics, an alternative
definition of $wlp.$ $(\while\ q$ $\ \ddo\ S).M$ is the greatest
fixed point $\nu X\cdot(\qzz wlp.S.X \qzz + \qoo$ $ M\qoo)$.

The following theorem shows a quantitative connection between
denotational semantics and weakest liberal precondition semantics.
\begin{thm}\label{thm:wlpop}
For any purely quantum program $S$, quantum predicate $M\in\ph$, and
$\rho\in\dh$, we have
\begin{equation}\label{eq:wlp}
\tr(wlp.S.M)\rho = \tr M\sem{S}\rho + \tr\rho - \tr\sem{S}\rho.
\end{equation}
\end{thm}
{\it Proof.} Similar to Theorem \ref{thm:wpop}.  \hfill $\square$

\vspace{1em}

Taking $M=\z$ in Eq.(\ref{eq:wlp}), we have $$\tr(wlp.S.\z)\rho =
\tr\rho -\tr\sem{S}\rho.$$ Notice that the righthand side of the
above equation denotes the probability the program $S$ does not
terminate when the input state is $\rho$. So intuitively the quantum
predicate $wlp.S.\z$ denotes the condition the program $S$ diverges.

\begin{cor}\label{cor:wpwlp} For any purely quantum program $S$ and quantum predicate $M\in\ph$,
$$wp.S.M\le wlp.S.M$$ and
$$wlp.S.M+wp.S.(I-M)=I$$
\end{cor}

\vspace{1em}

To get a clearer picture of the connection between these two
precondition semantics, let us introduce a notion which is the
analogue of conjunction $\wedge$ of classical standard predicates
and probabilistic conjunction $\&$ of classical probabilistic
predicates (see, for example, \cite{Mo95}).

\begin{defn} Suppose $M$ and $N$ are two quantum predicate. We define $M\& N$
as a new predicate
$$M\& N:= (M+N-I)^+,$$
where for any Hermitian matrix $X$, if $X=\sum_i \lambda_i P_i$ is
the spectrum decomposition of $X$, then $X^+=\sum_i \max
\{\lambda_i,0\}P_i$. It is obvious that if $M+N\ge I$, then $M\&
N=M+N-I$.
\end{defn}

\begin{thm}\label{thm:wand}
For any quantum predicates $M,N\in\ph$ and any purely quantum
program $S$, if $M+N\ge I$ then
\begin{equation}\label{eq:wpwlp} wp.S.(M\&N)=wp.S.M\ \&\ wlp.S.N
\end{equation} and
\begin{equation}\label{eq:wpwlp1}
wlp.S.(M\&N)=wlp.S.M\ \&\ wlp.S.N
\end{equation}
\end{thm}
{\it Proof.} We only prove Eq.(\ref{eq:wpwlp}); the proof of
Eq.(\ref{eq:wpwlp1}) is similar. From the assumption that $M+N\ge
I$, we have $M\& N=M+N-I$. Then for any $\rho : \dh$,
\begin{eqnarray*}
& &\tr wp.S.(M\& N)\rho\\
& =&\tr wp.S.(M+N- I) \rho \\
&=&\tr(M+N- I)\sem{S}\rho \hspace{10.6em}\mbox{Theorem \ref{thm:wpop}}\\
&=&\tr M\sem{S}\rho+\tr N\sem{S}\rho-\tr \sem{S}\rho\\
&=&\tr wp.S.M \rho+\tr wlp.S.N\rho-\tr\rho\hspace{5.3em}\mbox{Theorems \ref{thm:wpop} and \ref{thm:wlpop}}\\
&=&\tr(wp.S.M+wlp.S.N-I)\rho
\end{eqnarray*}
So we have $wp.S.(M\& N)= wp.S.M + wlp.S.N - I$ and then $wp.S.(M\&
N)= wp.S.M\ \&\ wlp.S.N$ from the fact that $wp.S.(M\& N)\ge \z$.
\hfill $\square$

\vspace{1em}

When taking $N=I$ in Eq.(\ref{eq:wpwlp}), we have the following
direct but useful corollary:

\begin{cor}\label{cor:andor} For any purely quantum program $S$ and quantum predicate
$M$,
\begin{equation}\label{eq:term}
wp.S.M=wp.S.I\ \&\ wlp.S.M
\end{equation}
\end{cor}

Recall that $wp.S.I$ denotes the condition the program $S$
terminates. So the intuitive meaning of Eq.(\ref{eq:term}) is that a
program is total correct (represented by weakest precondition
semantics) if and only if it is partial correct (represented by
weakest liberal precondition semantics) $and$ it terminates. This
capture exactly the intuition of total correctness and partial
correctness.

To conclude this section, we present some properties of weakest
liberal precondition semantics which are useful in the next section.
The proofs are direct so we omit the details here.
\begin{lem}\label{lem:wlppro} For any purely quantum program $S$ and quantum predicate
$M,N\in\ph$, we have

1) $wlp.S.I=I$;

2) (monotonicity)  if $M\le N$ then $wlp.S.M\le wlp.S.N$;

3) if $M+N\le I$ then $wlp.S.(M + N)=wp.S.M + wlp.S.N$;

4) if $M\ge N$ then $wlp.S.(M - N)=wlp.S.M - wp.S.N$.
\end{lem}

\section{Proof rules for quantum loops}
Proof rules for programs are important on the way to designing more
general refinement techniques for programming. In this section, we
derive some rules for reasoning about loops in our purely quantum
language fragment. We find that almost all loop rules derived in
classical probabilistic programming (see, for example, \cite{Mo95}
or \cite{MM04}) can be extended to quantum case.

In classical standard or probabilistic programming languages, an
appropriate invariant is the key to reasoning about loops. It is
also true in quantum case. So our first theorem is devote to
reasoning about quantum loops within partial correctness setting
using $wlp$-invariants. Recall that in classical probabilistic
programming, if $Inv$ is a $wlp$-invariant of a loop statement
$loop\equiv ``\while\ b\ \ddo\ S"$ satisfying
$$[b]*Inv \Rrightarrow wlp.S.Inv,$$ then
$$Inv\Rrightarrow wlp.loop.([\overline{b}]*Inv).$$ Here $\Rrightarrow$
means ``everywhere no more than", which is the probabilistic
analogue of the implication relation ``$\Rightarrow$" in standard
logic.

\begin{thm}\label{thm:wlp} For any quantum predicate $M\in\ph$, if \begin{equation}\label{loopinv}
|0\>\<0|M|0\>\<0|\le wlp.S.(\sum_{i=0}^1|i\>\<i|M|i\>\<i|)
\end{equation} then
$$\sum_{i=0}^1|i\>\<i|M|i\>\<i| \le
wlp.qloop.(|1\>\<1|M|1\>\<1|).$$ Here and in what follows, by
$qloop$ we denote the quantum program ``$\while\ q\ \ddo\ S$''.
\end{thm}
{\it Proof.} By definition, we have
$$wlp.qloop.(|1\>\<1|M|1\>\<1|)=\sqcap_{j=0}^\infty M_j,$$
where $M_0=I$ and for $j\geq 1$,
$$M_{j+1}=|0\>\<0|wlp.S.M_j|0\>\<0| + |1\>\<1|M|1\>\<1|.$$
In what follows, we prove by induction that for any $j\geq 0$,
\begin{equation}\label{mi}
\sum_{i=0}^1|i\>\<i|M|i\>\<i| \le M_j.\end{equation}

When $j=0$, Eq.(\ref{mi}) holds trivially. Suppose Eq.(\ref{mi})
holds for $j=k$. Then when $j=k+1$, we have
\begin{eqnarray*}
M_{k+1}&=&|0\>\<0|wlp.S.M_k|0\>\<0| + |1\>\<1|M|1\>\<1|\\ &\ge&
|0\>\<0|wlp.S.(\sum_{i=0}^1|i\>\<i|M|i\>\<i|)|0\>\<0| +
|1\>\<1|M|1\>\<1|\\
& & \hspace{13em}\mbox{induction hypothesis and Lemma
\ref{lem:wlppro} (2)}
\\&\ge & |0\>\<0|M|0\>\<0|+ |1\>\<1|M|1\>\<1|.
\hspace{15em}\mbox{Eq.(\ref{loopinv})}
\end{eqnarray*} With that we complete the proof of this theorem.
 \hfill $\square$

\vspace{1em}

We say $\sum_{i=0}^1 |i\>\<i|M|i\>\<i|$ is a $wlp$-invariant of
$qloop$ if Eq.(\ref{loopinv}) holds; similarly, $\sum_{i=0}^1
|i\>\<i|M|i\>\<i|$ is a $wp$-invariant of $qloop$ if
\begin{equation}\label{loopinv1} |0\>\<0|M|0\>\<0|\le
wp.S.(\sum_{i=0}^1|i\>\<i|M|i\>\<i|).
\end{equation}

We now turn to reasoning about quantum loops in total correctness
setting. Following the remark behind Corollary \ref{cor:andor}, we
give the total correctness of quantum loops by combining partial
correctness with the termination condition. To simplify notations,
we define
$$T:=wp.qloop.I.$$ Intuitively, $T$
denotes the termination condition of $qloop$.

\vspace{1em}

For any quantum loop, if a $wp$-invariant implies the termination
condition, then its partial correctness is sufficient to guarantee
its total correctness, as the following lemma states.

\begin{lem}\label{lem:wpsubT} For any quantum predicate $M\in\ph$, if $\sum_{i=0}^1 |i\>\<i|M|i\>\<i|$ is a $wp$-invariant of $qloop$,
$wp.S.T\le T$, and
\begin{equation}\label{assu}
\sum_{i=0}^1|i\>\<i|M|i\>\<i|\le T, \end{equation} then
$$\sum_{i=0}^1|i\>\<i|M|i\>\<i|\le
wp.qloop.(|1\>\<1|M|1\>\<1|).$$
\end{lem}
{\it Proof.} Let $$M':=\sum_{i=0}^1|i\>\<i|M|i\>\<i|+I-T.$$ Noticing
that from definition we have $T=|0\>\<0|wp.S.T|0\>\<0| + |1\>\<1|$,
then
\begin{equation}\label{t0}
|0\>\<0|T|0\>\<0|=|0\>\<0|wp.S.T|0\>\<0|,
\end{equation}
\begin{equation}\label{t1}
|1\>\<1|T|1\>\<1|=|1\>\<1|
\end{equation}
 and
\begin{equation}\label{t01}
\sum_{i=0}^1|i\>\<i|T|i\>\<i|=T.
\end{equation} Furthermore, we can check that
$M'=\sum_{i=0}^1|i\>\<i|M'|i\>\<i|$. From the assumption
Eq.(\ref{assu}), we can easily derive $\z\le$ $M'\le I$, and $M'$ is
also a quantum predicate on $\h$. We calculate
\begin{eqnarray*}
& &wlp.S.\sum|i\>\<i|M'|i\>\<i| \\
&=&wlp.S.(\sum|i\>\<i|M|i\>\<i|+I-T)\\
&=&wp.S.\sum|i\>\<i|M|i\>\<i|+wlp.S.(I-T) \hspace{7.7em}\mbox{Lemma }\ref{lem:wlppro}\ (3)\\
&=&wp.S.\sum|i\>\<i|M|i\>\<i|+wlp.S.I-wp.S.T\hspace{6em}\mbox{Lemma }\ref{lem:wlppro}\ (4)\\
&\ge &|0\>\<0|M|0\>\<0| + I - T \hspace{14.7em}\mbox{Lemma \ref{lem:wlppro}}\ (1) \mbox{ and Eq.}(\ref{loopinv1})\\
&= &|0\>\<0|M|0\>\<0| + |0\>\<0| - |0\>\<0|T|0\>\<0|
\hspace{8.2em}\mbox{Eqs.}(\ref{t0})-(\ref{t01})\\
&=&|0\>\<0|M'|0\>\<0|.
\end{eqnarray*} So $\sum_{i=0}^1|i\>\<i|M'|i\>\<i|$ is a $wlp$-invariant of $qloop$.
We further calculate

\begin{eqnarray*}
& & \sum_{i=0}^1|i\>\<i|M|i\>\<i|  \\
& =& M' + T -I  \hspace{19.4em} \mbox{definition of }M'\\
& \le & wlp.qloop.(|1\>\<1|M'|1\>\<1|)+T-I\hspace{9.4em} \mbox{Theorem }\ref{thm:wlp}\\
& = & wlp.qloop.(|1\>\<1|M'|1\>\<1|)\ \&\
T\hspace{11em}\mbox{Corollary
\ref{cor:wpwlp}}\\
& = & wp.qloop.(|1\>\<1|M'|1\>\<1|) \hspace{13.5em}\mbox{Eq.
(\ref{eq:term})}\\
& =& wp.qloop.(|1\>\<1|M|1\>\<1|). \hspace{13.5em} \mbox{Eq.
}(\ref{t1})
 \end{eqnarray*}
That completes our proof.
 \hfill $\square$

\vspace{1em}

To conclude this section, we generalize the powerful 0-1 law in
classical programming to quantum case. Informally, 0-1 law states
that if the probability of a loop terminating from a state is at
least $p$ for some fixed $p>0$ (no matter how small $p$ is), then
the loop terminates with certainty when started from that state. In
other words, the terminating probability is either 0 or 1 and cannot
lie properly between 0 and 1.

\begin{lem}\label{lem:smallp} For any quantum predicate $M\in\ph$, if
$\sum_{i=0}^1|i\>\<i|M|i\>\<i|$ is a $wp$-invariant of $qloop$,
$wp.S.T\le T$, and $\exists\ 0<p\leq 1$ such that
$p*\sum_{i=0}^1|i\>\<i|M|i\>\<i|\le T$ then
$$\sum_{i=0}^1|i\>\<i|M|i\>\<i|\le T.$$
\end{lem}
{\it Proof.} Let $M':=p*M$. Then $\sum_{i=0}^1|i\>\<i|M'|i\>\<i|\le
T$ and furthermore,
\begin{eqnarray*}
|0\>\<0|M'|0\>\<0|&=&
p*|0\>\<0|M|0\>\<0|\\
&\le&p*wp.S.(\sum_{i=0}^1|i\>\<i|M|i\>\<i|)\\
&=&wp.S.(\sum_{i=0}^1|i\>\<i|M'|i\>\<i|).\hspace{3.8em}\mbox{linearity
of } wp.S
\end{eqnarray*}
So we can derive that
\begin{eqnarray*}
p*\sum_{i=0}^1|i\>\<i|M|i\>\<i|&=&\sum_{i=0}^1|i\>\<i|M'|i\>\<i|\\
&\le& wp.qloop.(|1\>\<1|M'|1\>\<1|)\hspace{3em}\mbox{Lemma \ref{lem:wpsubT}}\\
&=& p*wp.qloop.(|1\>\<1|M|1\>\<1|)\hspace{1.8em}\mbox{linearity of }wp.qloop\\
&\le & p*wp.qloop.I\hspace{7.8em}\mbox{monotonicity of }wp.qloop\\
&= & p*T.
\end{eqnarray*}
Dividing both sides by the positive number $p$, we arrive at the
desired result.
 \hfill $\square$
\begin{thm}(0-1 law for quantum loops) If $\ T$ is
positive-definite and $wp.S.T\le T$, then for any quantum predicate
$M\in\ph$ such that
$$|0\>\<0|M|0\>\<0|\le
wp.S.(\sum_{i=0}^1|i\>\<i|M|i\>\<i|),$$ we have
$$\sum_{i=0}^1|i\>\<i|M|i\>\<i| \le
wp.qloop.(|1\>\<1|M|1\>\<1|).$$
\end{thm}
{\it Proof.} From the assumption that $T$ is positive-definite, for
any $wp$-invariant $\sum_{i=0}^1|i\>\<i|M|i\>\<i|$ of $qloop$ there
exists a sufficiently small but positive $p$ such that
$p*\sum_{i=0}^1|i\>\<i|M|i\>\<i|$ $\le T$. So
$\sum_{i=0}^1|i\>\<i|M|i\>\<i|\le T$ from Lemma \ref{lem:smallp}.
Then the result of this theorem holds by applying Lemma
\ref{lem:wpsubT}.
 \hfill $\square$

\section{Conclusion}
In this paper, we applied the notion of quantum predicate proposed
by D'Hondt and Panangaden in \cite{HP05} to analyze a purely quantum
language fragment which involves only quantum-typed variables. This
language can be treated as the quantum fragment of a general
programming language, describing the quantum device of a future
quantum computer in Knill's architecture. The denotational semantics
of this language was introduced. We further proposed the weakest
precondition semantics and weakest liberal precondition semantics,
corresponding respectively to total and partial correctness of
quantum programs. The connections between these three semantics were
discussed. To help reasoning about quantum loops, we extended all
existing proof rules for loops in classical probabilistic programs
to the case of our purely quantum programs.

\section*{Acknowledgement}
The authors thank the colleagues in the Quantum Computation and
Quantum Information Research Group for useful discussion. This work
was partly supported by the Natural Science Foundation of China
(Grant Nos. 60503001, 60321002, and 60305005), and by Tsinghua Basic
Research Foundation (Grant No. 052220204). R. Duan acknowledges the
financial support of Tsinghua University (Grant No. 052420003).

\bibliographystyle{plain}

\end{document}